**Divergent Coherent Phonon Responses Across the Metal-Insulator Crossover**

*Felix Hoff, Timo Veslin, Tim Bartsch, Carl-Friedrich Schön, Dante M. Kennes, Matthias Wuttig*

F. Hoff, T. Veslin, T. Bartsch, C-F. Schön, M. Wuttig

Institute of Physics IA, RWTH Aachen University, 52074 Aachen, Germany

E-Mail: wuttig@physik.rwth-aachen.de

D. M. Kennes

Institute for Theory of Statistical Physics, RWTH Aachen University, and JARA Fundamentals of Future Information Technology, 52074 Aachen, Germany

D. M. Kennes

Max Planck Institute for the Structure and Dynamics of Matter, 22761 Hamburg, Germany

M. Wuttig

Peter-Grünberg-Institute – JARA-Institute Energy Efficient Information Technology (PGI-10), 52428 Jülich, Germany

**Abstract**

Ultrafast laser control of material properties hinges on understanding light-matter interactions. We use two experimentally accessible response functions, laser fluence induced phonon softening and the amplitude of coherent reflectance oscillations, to compare how strongly different materials respond to ultrafast photoexcitation. Comparing a diverse set of materials, we find that only a narrow class, including Sb, GeTe, and $Bi_2Te_3$, shows exceptional responses such as pronounced phonon softening and a giant increase of reflectance oscillations with increasing fluence. These response functions peak in an intermediate conductivity regime of about $10^2$ - $10^4$ S/cm, at the crossover between localized and delocalized electronic states. The corresponding class of solids also shows other unconventional properties including high dielectric constants, enhanced Born effective charges, coordination numbers exceeding the 8-N rule and uncommon bond rupture. This suggests that these materials employ a unique bonding mechanism, coined metavalent bonding. Frozen-phonon DFT calculations show that the strong fluence dependence arises from Peierls-like instabilities, leading to large deformation potentials and anharmonic double-well potentials. These findings identify metavalent bonding as a design principle for enhanced coherent phonon control and provide a quantitative framework for identifying materials with exceptional ultrafast responses.

## Introduction

Ultrafast light-matter interaction has emerged as a powerful approach to probe and control the macroscopic properties of quantum materials on femtosecond timescales [1, 2]. Using intense, ultrashort laser pulses, it becomes possible not only to observe the rapid dynamics of electrons and the lattice, but also to drive materials into novel non-equilibrium states with tailored functionalities [3-5]. Among the insightful phenomena accessible in this regime are coherent phonons, phase-locked atomic oscillations of the crystal lattice that are launched by optical excitation [6, 7]. These collective vibrations directly link atomic motion to changes in electronic structure, enabling the real-time investigation of fundamental interactions between electrons and lattice. As such, coherent phonons serve as sensitive probes for studying phase transitions, competing quantum ground states, and emergent phenomena in complex solids [8-10].

Coherent phonons have been extensively investigated using advanced time-resolved techniques such as time-resolved x-ray diffraction (tr-XRD), [1, 11] time- and angle-resolved photoemission spectroscopy (tr-ARPES),[12] transient optical spectroscopy (TOS),[6] and ultrafast electron diffraction (UED).[13] Many state-of-the-art studies focus on single quantum materials, exploring how coherent lattice dynamics respond to variations in excitation density,[1, 2, 7] ambient temperature,[2, 14] or pressure.[15] These approaches have provided valuable insights into the microscopic mechanisms governing phase transitions, including charge-density wave (CDW) formation and metal-to-insulator transitions.[9, 14, 16] However, systematic comparisons of coherent phonons as a response function across different material classes remain rare.

The observation that certain materials exhibit much stronger coherent phonon responses than others was already highlighted in landmark work on displacive excitation of coherent phonons (DECP) in the early 1990s. [6, 17] In this context, it was noted that oscillation amplitudes in reflectivity for select materials can be orders of magnitude larger compared to typical metals, with only specific symmetry modes (i.e. $A_{1g}$) being efficiently excited. [6, 17] However, it remains unclear which materials host such exceptionally strong coherent phonon responses and what microscopic mechanisms give rise to them. Identifying these materials and understanding the origin of their enhanced lattice dynamics are key questions addressed in this study.

Coherent phonon generation in absorbing solids can be described within the framework of stimulated Raman processes, which encompasses both the impulsive Raman limit and the displacive excitation mechanism. [18, 19] In this picture, the oscillatory reflectivity signal measured in pump-probe experiments reflects not only the efficiency with which the pump pulse launches the lattice coordinate but also the susceptibility of the phonon mode and the sensitivity with which the probe detects the resulting modulation of the dielectric function. Consequently, the measured coherent phonon amplitude should be interpreted as a combined generation-and-detection response rather than as a direct measure of a bare electron-phonon coupling constant. [18, 19]

Several of the materials considered in this work have already been investigated individually in the high-excitation regime. Excitation-density-dependent redshifts of coherent optical phonons have been reported, for example, in elemental semimetals such as Bi, [20] Sb, [21] and Te, [22] in phase-change related materials such as GeTe, [23, 24] and in topological compounds including $Bi_2Te_3$. [25, 26] These studies demonstrated the pronounced sensitivity to ultrafast optical excitation for selected semimetals and phase-change-related materials. However, a systematic, quantitative comparison of coherent phonon response functions across different material classes under consistent experimental conditions has so far been lacking.  Consequently, the questions raised above, how to identify the class of materials that exhibits exceptionally strong coherent phonon responses and which material properties govern this behavior, have remained largely unresolved.

Here, we present a comparative study of coherent phonons spanning covalent semiconductors, metavalent semiconductors/semimetals, topological chiral semimetals, [27] and metals. We employ an all-optical femtosecond pump-probe setup, extracting both the amplitude-increase and frequency softening of coherent phonons as a function of absorbed pump fluence. By using the absorbed fluence, which incorporates the sample thickness in conjunction with the dielectric function and provides a consistent first-order normalization of the deposited optical energy, we ensure a meaningful comparison across materials. Transient reflectance measurements provide access to phonon generation and detection with high temporal resolution, as shown schematically in **Figure 1**a.

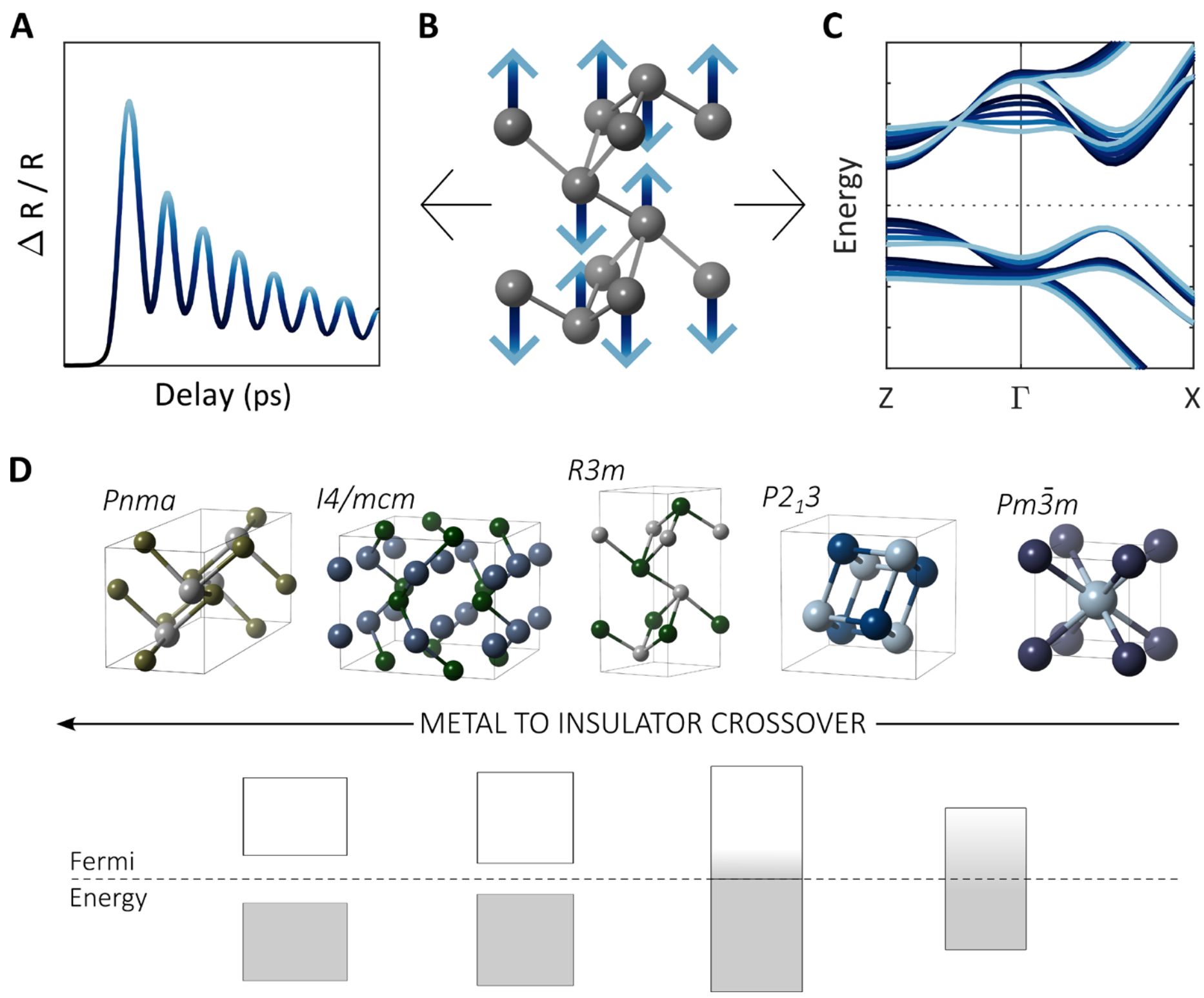

**Figure 1**: (a) Representative transient reflectivity measurement showing coherent optical phonon dynamics on the picosecond timescale. (b) Conventional unit cell of the material with phonon displacement pattern indicated by arrows; arrow color gradient encodes atomic displacement amplitude. (c) Schematic electron band structure illustrating changes induced by varying amplitudes of atomic displacement along the wave vector of the phonon mode, with corresponding color coding for increasing phonon amplitude. (d) Overview of some representative crystal structures and schematic electronic density of states (DOS) motifs characteristic of the material classes studied to cross from metals to insulators. Shown are schematic unit cells and corresponding DOS sketches near the Fermi level, illustrating typical structural symmetries and electronic properties for insulators, semiconductors, semimetals, and metals included in this work.

Within this framework, the excitation-dependent increase of coherent phonon amplitude and frequency softening serve as quantitative experimental metrics for classifying materials across diverse bonding regimes, especially near the transition between localized and delocalized charge carriers. By coupling ultrafast optical measurements with first-principles frozen phonon calculations (Figure 1c) and detailed analysis of orbital overlap effects, we establish a methodology to explore electron-lattice interactions in systems with competing ground states. The variety of crystal structures, band gaps and electronic density of states investigated in this study is schematically illustrated in Figure 1d. This approach provides quantitative benchmarks for comparing the strength of coherent phonon generation and fluence-dependent phonon softening across materials with very different bonding regimes. In particular, our results identify a distinct class of solids as exceptionally sensitive to ultrafast excitation. These observations are rationalized using frozen-phonon calculations and an analysis of orbital-overlap effects, providing microscopic insight into why specific phonon coordinates in these materials respond strongly to optical excitation.

**Design of Comparative Ultrafast Experiments**

Time-resolved optical spectroscopy with a two-color femtosecond pump-probe setup was employed to measure coherent phonon dynamics across a considerable range of materials. The experiments were performed with an 800 nm pump pulse of 60 fs duration and a 516 nm probe pulse of 80 fs duration. For quantitative comparison, the measured response functions, such as the increase of amplitude and phonon frequency softening, are evaluated as a function of absorbed pump fluence. This approach accounts for differences in sample thickness, optical constants, and reflectance. It ensures that the excitation density is directly related to the number of photoexcited carriers in each material, enabling a meaningful comparison across diverse material classes.

With these excitation conditions established, we analyze how ultrafast optical stimulation drives lattice and electronic dynamics on the microscopic scale. Upon photoexcitation, a system can be described by a two-temperature model, where electrons and the lattice are treated as coupled but distinct subsystems. [28, 29] The pump pulse rapidly increases the electronic temperature within the duration of the pump pulse, generating electron-hole pairs and inducing an initial change in reflectance due to modifications of the material’s dielectric function. [30] As energy is transferred from the hot electrons to the lattice, the reflectivity signal begins to recover, indicating electronic cooling.  In materials with distorted lattice structures, such as those exhibiting Peierls-like distortions or charge density waves (CDWs), the elevated electron temperature can drive atoms collectively toward a new equilibrium position with reduced distortion, a process known as DECP. Upon laser-excitation antibonding states above the Fermi energy are occupied which reduces the size of the Peierls-like distortion. In other crystals, coherent optical phonons may be launched primarily through impulsive stimulated Raman scattering (ISRS), which involves direct transfer of momentum from the ultrafast pulse to the vibrational modes without requiring a change in equilibrium atomic positions. In both cases, coherent oscillations manifest as periodic modulations in the transient reflectivity signal, providing direct insight into both ultrafast lattice dynamics and their coupling to electronic states.[6, 17]

**Material-Dependent Amplitude and Softening of Coherent Phonons**

We investigated a set of solids spanning more than fourteen orders of magnitude in room temperature conductivity, from $10^{-8}$ S/cm to $10^{6}$ S/cm. Covalent solids such as GeSe ($Pnma$ structure) and InAs (zincblende structure), where bonding electrons are localized between ion cores, exhibit low conductivities below $10^{2}$ S/cm. In contrast, incipient metals including antimony ($R\overline{3}m$ bilayer structure) and GeTe ($R3m$ structure), [31] as well as elemental tellurium ($P3_{1}21$ structure), fall into an intermediate regime with conductivities between $10^{2}$ and $10^{4}$ S/cm. Finally, intermetallic compounds such as AlPt ($P2_{1}3$, B20 chiral structure) exhibit higher conductivities associated with more metallic bonding. The broad range of conductivities was chosen to probe how coherent phonon responses evolve across different electronic regimes. In particular, several unconventional material properties are known to emerge in the intermediate conductivity range around $10^{2}$ to $10^{4}$ S/cm, where electronic localization and delocalization compete. [31-34]

**Figure 2** presents representative transient reflectance traces for the six materials mentioned above at their respective highest reversible absorbed fluences that still yields fully reversible lattice dynamics, i.e., just below the onset of permanent sample modification. Clear coherent phonon oscillations are observed in all compounds, but their amplitudes vary strongly between materials. In particular, the incipient metals (Sb, GeTe, and Te) display markedly larger oscillation amplitudes than the covalent semiconductors (GeSe, InAs) and the chiral semimetal AlPt. For comparison, we note that coherent optical phonons have previously also been observed in simple metals such as Zn and Cd using ultrafast pump-probe spectroscopy, although their amplitudes are typically small and strongly damped at room temperature due to coupling to the electronic continuum. [35, 36]

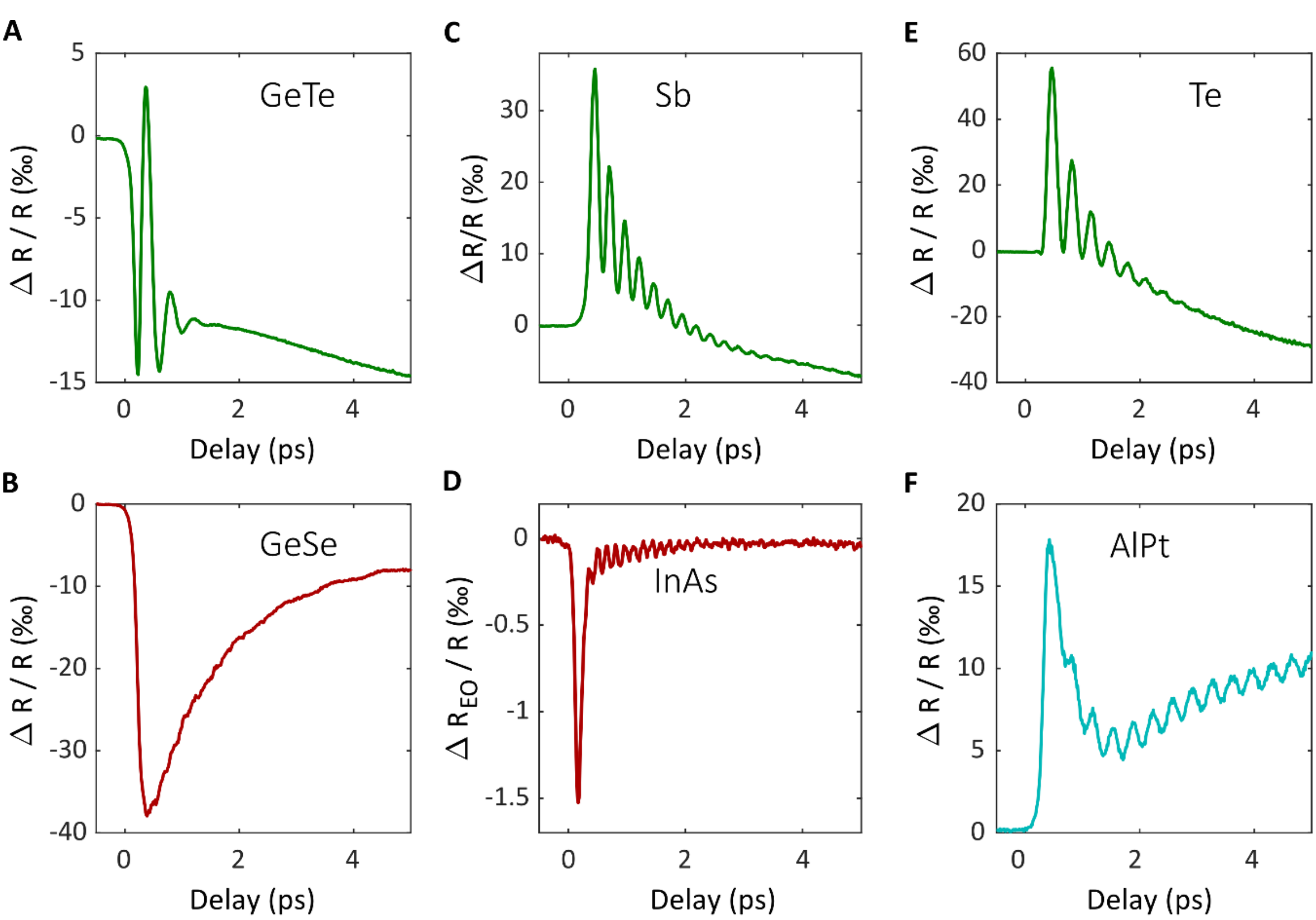

**Figure 2**: Representative transient reflectivity traces for six materials spanning distinct structural and electronic classes: GeTe (metavalent, $R3m$ structure), GeSe (covalent, $Pnma$ structure), Sb (metavalent, $R\overline{3}m$ structure), InAs (covalent, zincblende structure), Te (metavalent, helical chain structure, $P3_121$), and AlPt (chiral semimetal, B20 structure). All measurements were performed at the highest reversible excitation fluence just below the modification threshold. Isotropic $\Delta R/R$ shown for all except InAs, where the dominant TO phonon requires electro-optical detection (EO sampling). For clarity, each trace is displayed with an individual vertical scale chosen to highlight the oscillation amplitude.

To systematically compare coherent phonon responses across material classes, we analyze how the oscillation amplitude evolves as a function of absorbed pump fluence. The increase of the amplitude

$$S_{\mathrm{A}} = \partial A_{osc} / \partial F_{abs},$$

extracted from the slope of amplitude versus fluence, serves as a figure of merit for the coherent phonon response of each material. In general, the oscillatory reflectivity amplitude measured in pump-probe experiments depends on several factors, including the efficiency with which the pump pulse drives the lattice coordinate, the susceptibility of the reflectance to the phonon mode, and the efficiency with which the probe detects the resulting modulation of the dielectric function. In a simplified picture the oscillatory signal can be written as

$$A_{\mathrm{osc}}(\omega_{\mathrm{probe}}) \propto (\partial R/\,\partial Q)_{\omega_{\mathrm{probe}}} Q_0,$$

where $Q_0$ denotes the coherent phonon amplitude. Consequently, the amplitude slope used here should be interpreted as a mode-specific coherent phonon response coefficient. The same renormalization was also chosen for the fluence-induced phonon softening, facilitating a direct comparison of different materials. A statistical analysis based on repeated measurements and weighted linear regression confirms that the uncertainties of the extracted slopes are significantly smaller than the orders-of-magnitude variation observed across different material classes (see Supplementary Information). The slopes were obtained within the reversible low-to-intermediate excitation regime of each material, i.e., below the onset of amplitude saturation, strong frequency chirping, or irreversible sample modification. While several complementary methods exist for quantifying oscillation strength, including time-domain and frequency-domain metrics, as well as fit-based approaches, all yield highly correlated results (see Supplementary Information). Importantly, the trends observed for the increase of amplitude are mirrored in the maximum measured amplitude for each material, confirming the consistency and reliability of our analysis.

**Figure 3**a presents the amplitude increase for each material as a function of room temperature electrical conductivity. The data span many orders of magnitude in conductivity, capturing covalent semiconductors at low conductivities ($< 10^2$ S/cm), good metals above $10^4$ S/cm, and metavalent solids as well as topological chiral semimetals in the intermediate regime ($10^2$– $10^4$ S/cm). Materials are color-coded according to a multi-property bonding classification scheme: covalent materials (red), metavalent solids (green), and topological chiral semimetals (cyan), based on bond-breaking behavior, optical properties, bond polarizability, effective coordination number, and related metrics. [31, 37, 38]

A striking divergence is observed: while covalent semiconductors, and topological chiral semimetals exhibit relatively modest responses, materials such as antimony and $Bi_2Te_3$, on the contrary, display significantly larger amplitude increase $S_A$. This emergent behavior suggests that these materials reside in a regime where certain phonon modes exhibit an unusually strong coherent reflectivity response under ultrafast excitation, resulting in enhanced oscillatory signals in the optical reflectivity traces. A summary of the investigated materials, their crystal structures, dominant phonon modes, and bonding assignments is provided in Table S1 in the Supplementary Information.

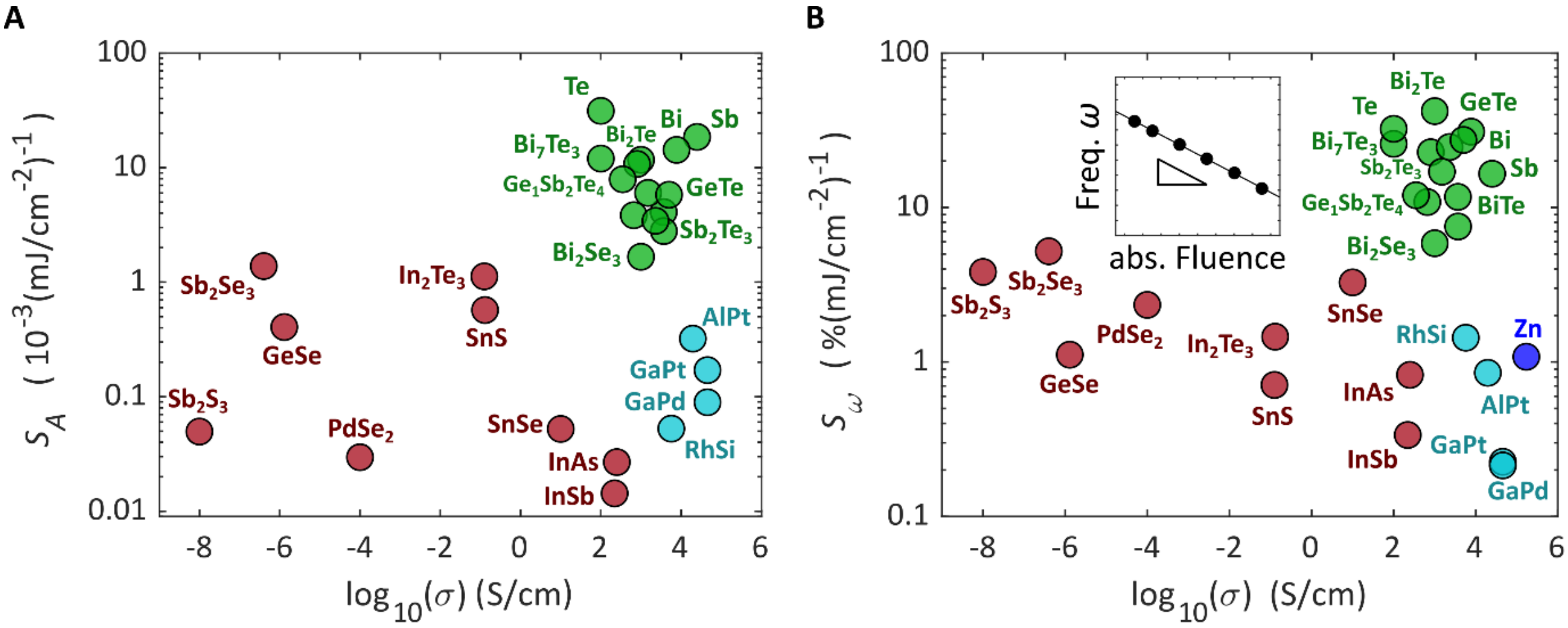

**Figure 3**: (**A**) Amplitude increase ($S_A$, in per mille per (mJ/cm²)) of the coherent phonon response as a function of room temperature conductivity, plotted on a logarithmic axes. This metric quantifies the sensitivity of the lattice response to excitation density across material classes. Markers are color-coded by bonding classification: red (covalent solids), green (metavalent solids), cyan (topological chiral semimetals), based on bond-breaking behavior, optical properties, bond polarizability, and effective coordination. [31, 37] (**B**) Frequency softening rate ($S_\omega$, in % per (mJ/cm²)) as a function of room temperature conductivity, plotted on a logarithmic axes. Same color coding as in (a). Covalent solids and topological chiral semimetals show moderate or low softening rates (0.1-5 % per (mJ/cm²)), while metavalent incipient metals exhibit much higher rates (5-40 % per (mJ/cm²)). The inset illustrates a representative linear fit of phonon frequency versus absorbed fluence, from which the softening rate is determined. The blue data point corresponds to Zn and is extracted from literature measurements at 150 K, [36] included as a representative metallic reference.

To further elucidate the mechanisms behind the observed trends, we analyzed how coherent phonon frequencies evolve with increasing excitation density across the same set of materials (Figure 3b). Phonon softening was quantified by tracking the frequency shift

$$S_\omega = -\partial / \partial F_{abs}\, \omega/\omega_0$$

as a function of absorbed pump fluence. Absorbed fluence was calculated using the Beer-Lambert law, accounting for sample thickness, absorption coefficient, and surface reflectance. The extinction coefficient was obtained from dielectric function values at the pump wavelength, ensuring that our metric reflects the true excitation density within each material rather than just the incident fluence. For each sample, a linear fit of phonon frequency versus absorbed fluence (as shown in the inset Figure 3b) provides a softening rate, expressed as a percentage change per unit fluence.

This analysis reveals that the same materials, such as Sb, GeTe, and $Bi_2Te_3$, exhibit both pronounced phonon softening and giant coherent phonon amplitude increase with growing excitation density, while covalent semiconductors and topological chiral semimetals display much weaker or negligible effects in both response functions.

In summary, the amplitude-increase $S_{\mathrm{A}}$ and phonon softening rate $S_{\omega}$ emerge as two complementary response functions, providing robust and quantitative metrics for comparing ultrafast lattice dynamics across diverse material classes. The pronounced response of metavalent solids highlights a unique sensitivity to photoexcitation in these materials.

**Insights from Frozen Phonon Calculations**

To rationalize the material-dependent coherent phonon amplitudes, we performed first-principles frozen phonon DFT calculations for each compound focusing on the most strongly visible zone-center phonon mode. Atomic displacements were chosen on the scale of the quantum zero-point motion of the mode, ensuring physically meaningful amplitudes that naturally account for differences in mass and frequency.

From these frozen-phonon calculations, we analyzed how the electronic band structure evolves with lattice displacement along the relevant phonon coordinate. In particular, we examine the shifts of electronic states near the band edges or near the Fermi level along representative high-symmetry k-paths. These band shifts provide a qualitative measure of the deformation potentials associated with the phonon coordinate and allow us to compare the sensitivity of the electronic structure to lattice motion across different materials. The total-energy curves as a function of displacement reveal the corresponding lattice potential energy surface (PES), enabling an assessment of the anharmonicity through deviations from a purely harmonic potential or the emergence of double-well character.

Representative results for six systems spanning insulators to (semi-)metals (GeTe, GeSe, Te, InAs, Bi, AlPt) reveal two contrasting behaviors. In conventional insulators (GeSe, InAs) and (semi-)metals such as AlPt, the relevant zone-center optical phonon modes largely preserve local coordination and key bond lengths. In GeSe, the $A_g$ phonon mainly involves interlayer shear motions that weakly perturb the short covalent Ge-Se bonds. In zincblende InAs, the $T_2$ optical phonon modulates the bond length of the tetrahedral In-As network but leaves the overall bonding configuration largely intact, leading to only modest shifts of the electronic bands near the band edges. Similarly, in AlPt the A modes corresponds to a nearly rigid sublattice rotation that only weakly perturbs the electronic states near the Fermi level. As a result, the corresponding band structures change only modestly with lattice displacement and the lattice potentials remain close to harmonic.

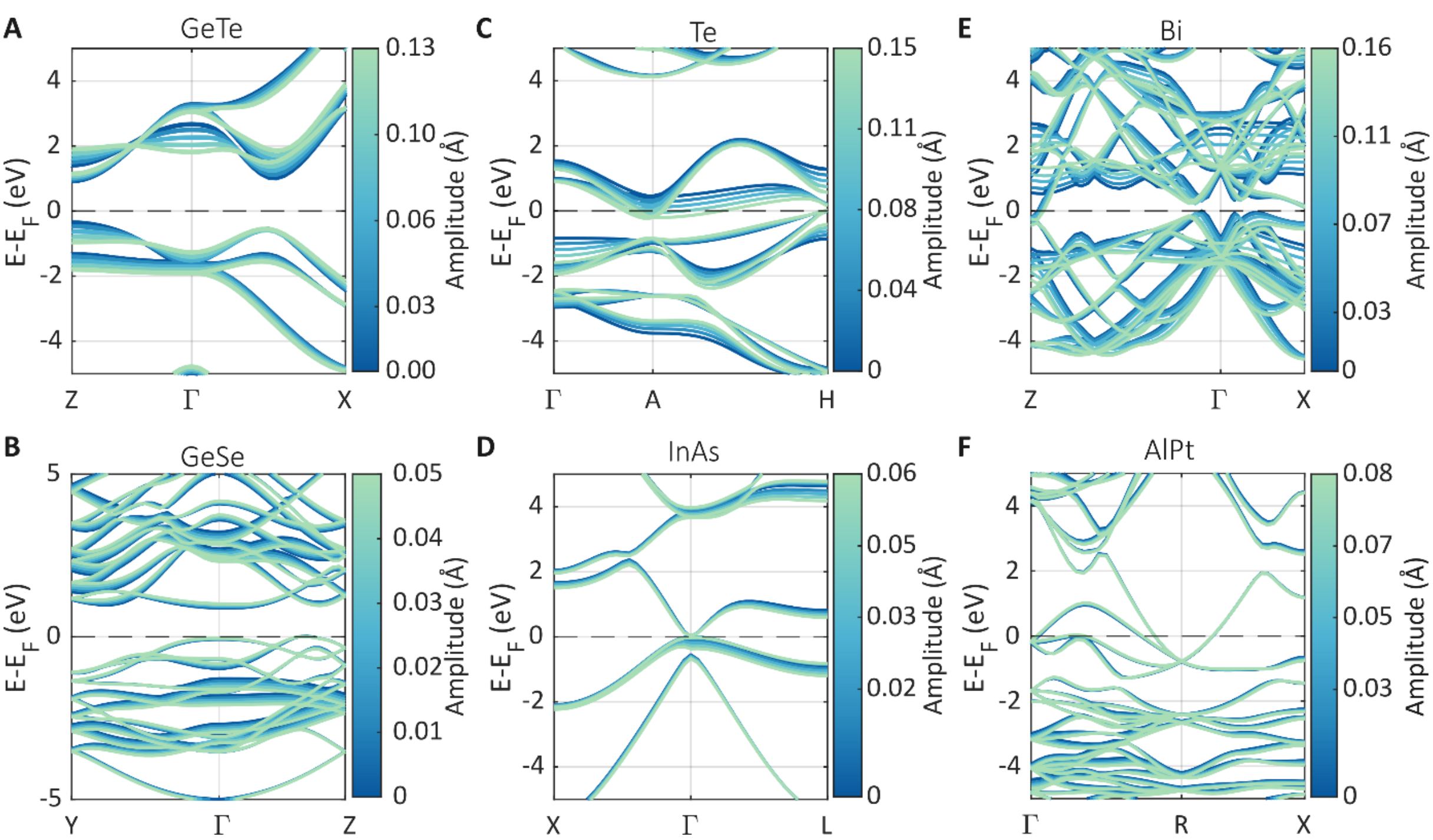

**Figure 4**: Frozen phonon band structures for six representative compounds, each shown for the strongest A-symmetry phonon mode observed experimentally. The energy window spans -5 to +5 eV around the Fermi level. The band structures are plotted along a k-path with three high-symmetry points including Γ; the color bar indicates atomic displacement amplitude in Å. The full band structures covering more extensive k-paths are provided in the Supplementary Information. Materials shown: (a) GeTe (*R3m*), (b) GeSe (*Pnma*), (c) Te ($P3_121$), (d) InAs (zincblende), (e) Bi (*R-3m*), and (f) AlPt ($P2_13$).

By contrast, in metavalent systems (GeTe, Te, Bi), the same *A* modes strongly modulate directional *p*-bond networks near structural instabilities. In GeTe, the $A_1$ mode tunes the Peierls-like dimerization along the trigonal axis. In Te, it unwinds the helical p-bonded chains; in Bi, it modulates the weakly dimerized bilayers. This produces large band-edge shifts (strong deformation potentials) and pronounced anharmonicity or double-well character in the total energy versus displacement curves.

The band structure changes induced by frozen phonon distortions directly affect the optical properties of the materials. To quantify this effect, we calculated several optical fingerprints as a function of the frozen phonon amplitude: the high-frequency dielectric constant $\varepsilon_\infty$, the maximum of the imaginary part of the dielectric function $\varepsilon_2$, and the Born effective charge $Z^*$. We have employed similar quantities previously as optical fingerprints for classifying bonding mechanisms in solids, where unusually large dielectric constants and Born effective charges signal the highly polarizable bonding characteristic of metavalent materials (incipient metals). [31] These quantities are also closely related to the optical reflectivity and lattice polarization that govern coherent phonon detection in pump-probe experiments.

**Figure 5** compares these optical fingerprints for a representative metavalent solid (GeTe) and a more conventional covalent semiconductor (GeSe). In GeTe, all three quantities, $\varepsilon_\infty$, the maximum of $\varepsilon_2$, and the Born effective charge $Z^*$, change strongly as the frozen phonon amplitude increases. This pronounced variation reflects the high polarizability of the metavalent bonding network, where small lattice distortions strongly modify p-orbital overlap and the associated electronic structure. By contrast, GeSe shows only weak changes of the same quantities over the same displacement range, consistent with its more rigid covalent bonding framework and the smaller deformation potentials discussed above. These results demonstrate that lattice distortions in metavalent materials simultaneously modulate both the electronic structure and the optical response, providing a direct microscopic link between the frozen phonon calculations and the large reflectivity oscillations observed experimentally.

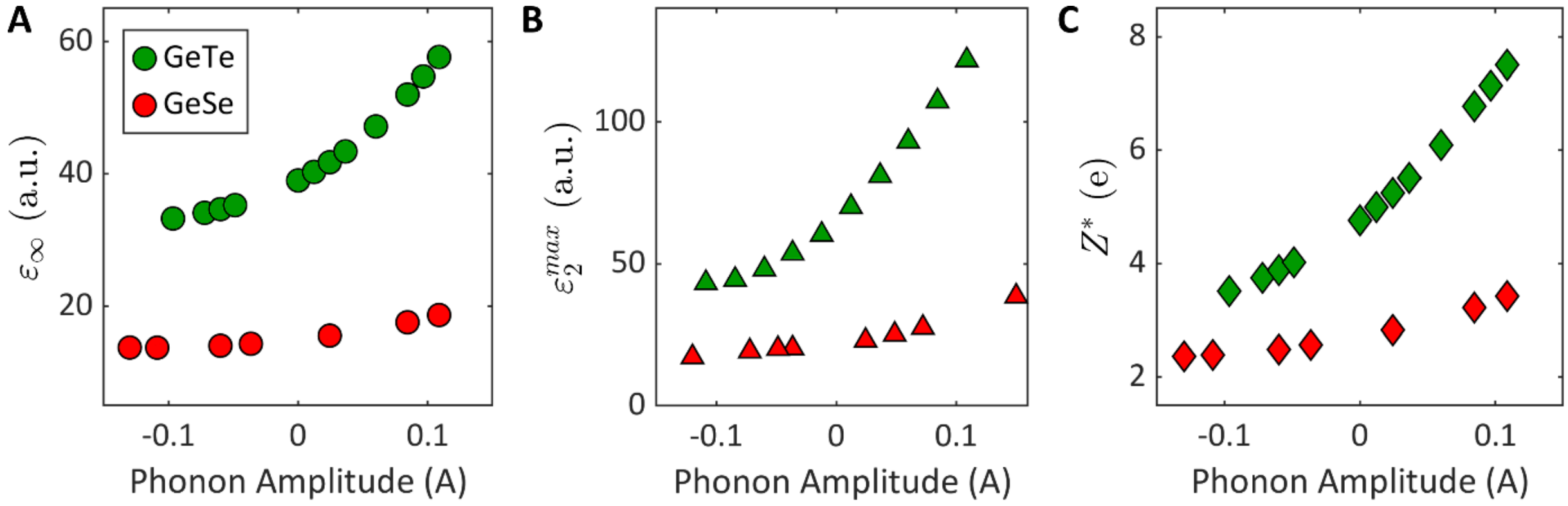

Figure 5: Optical fingerprints as a function of frozen phonon amplitude for GeTe (metavalent) and GeSe (covalent). (A) High-frequency dielectric constant $\varepsilon_{\infty}$, (B) Maximum value of $\varepsilon_2$, (C) Born effective charge $Z^*$. The first two quantities are a measure of the electronic polarizability and the strength of the optical absorption, respectively. The last quantity ($Z^*$), is a measure of the chemical bond polarizability and is closely related to the strength of electron-phonon coupling. The strong variation of all quantities in GeTe reflects the high polarizability of the metavalent bonding network, whereas GeSe shows only weak changes.

These results demonstrate that the same lattice distortion that drives the coherent phonon oscillation also significantly modifies the opto-electronic properties in metavalent materials. This provides a direct microscopic explanation for the large reflectivity oscillations observed experimentally. To understand why such small lattice distortions produce such large electronic and optical changes, we next examine the underlying lattice potential energy surfaces along the same phonon coordinates.

**Figure 6** summarizes the contrasting lattice potentials. The corresponding phonon displacement patterns for each compound are shown above the potential curves, highlighting the distinct atomic motions that define the experimentally observed coherent modes. Most materials (GeSe, AlPt, InAs) exhibit lattice potentials that remain close to parabolic near equilibrium along their dominant phonon mode coordinates over displacements of several pm, indicating comparatively weak anharmonicity typical of stable covalent (or metallic) bonding. By contrast, Sb, GeTe, and Te show strongly anharmonic profiles approaching double-well character. To quantify the anharmonicity of the lattice potential, the frozen-phonon energy curves were fitted with a polynomial expansion

$$V(Q) = \frac{1}{2}KQ^2 + \frac{1}{3}aQ^3 + \frac{1}{4}bQ^4.$$

The ratio $\eta = a/K$ in ($\text{Å}^{-1}$) provides a simple measure of the asymmetry and anharmonicity of the potential and is indicated for each material in Fig. 6. These Peierls-like distortions represent the competition between electron localization (dimerization or bond alternation) and delocalization (uniform metallic bonding): the phonon coordinate continuously tunes the balance between Peierls-like distorted (gapped or insulating-like) and undistorted (metallic) configurations. The double-well potentials reflect proximity to a structural instability where small electronic perturbations can trigger large lattice responses, explaining both the giant coherent amplitudes and strong fluence-dependent softening observed experimentally.

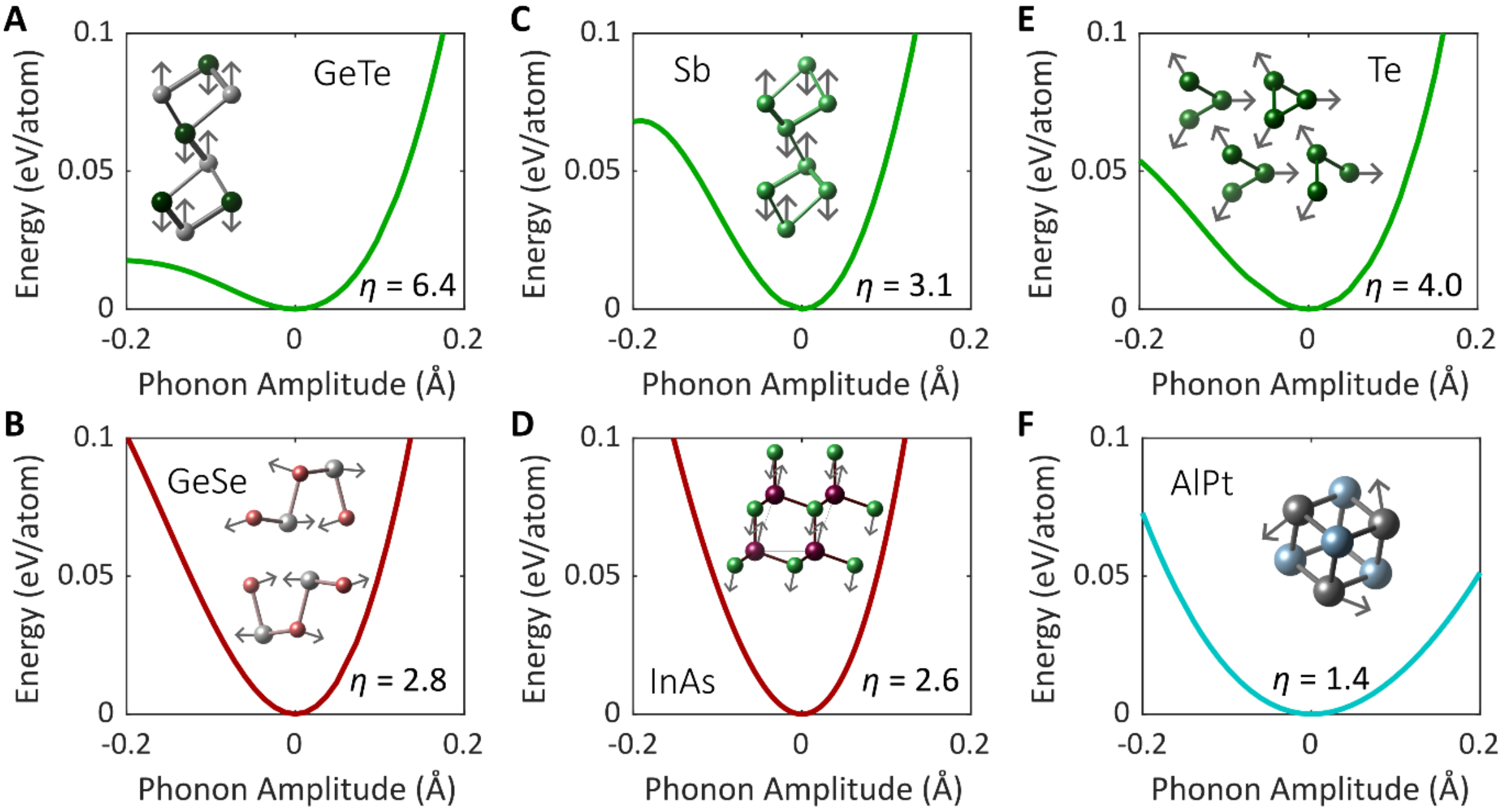

**Figure 6**: Lattice potential energy surfaces obtained from frozen-phonon calculations along the experimentally observed coherent phonon coordinates. The curves are color-coded according to the bonding classification introduced in Fig. 3. Insets show the corresponding phonon displacement patterns within the unit cell, with arrows indicating atomic motion. Metavalent systems (Sb, GeTe, and Te) exhibit strongly anharmonic, nearly double-well potentials characteristic of Peierls-distorted, metavalent bonding environments. In contrast, covalent semiconductors and more metallic systems (GeSe, InAs, and AlPt) display substantially weaker anharmonicity and potentials that remain close to harmonic near equilibrium. The ratio $\eta$ in $(\text{Å}^{-1})$, obtained from polynomial fits to the energy curves and indicated in each panel, provides a simple quantitative measure of the anharmonicity. Displacements up to ±0.2 Å are shown to illustrate the overall shape of the potentials energy surfaces and to highlight the emergence of anharmonicity and double-well character.

The strongly anharmonic double-well potentials and large deformation potentials in Sb, GeTe, and Te are consistent with metavalent bonding characteristics such as high mode-specific Grüneisen parameters and bond polarizability. However, our study probes a unique intermediate excitation regime, not employed before for material classification, where pump fluences of 0.5-10 mJ/cm² (peak fields of 0.05-0.15 MV/cm) drive displacive coherent phonons in an intermediate excitation regime stronger than linear-response Raman but weaker than nonthermal phase transitions. This regime reveals emergent nonlinear electron-lattice coupling not captured by equilibrium metavalent property measurements.

In summary, our combined frozen phonon and electronic structure analysis provides microscopic insight into the origin of the two experimentally observed response functions. The amplitude-increase $S_{\mathrm{A}}$ reflects the sensitivity of the electronic structure to lattice motion, captured by large band-edge deformation potentials, while the phonon softening rate $S_{\omega}$ reflects the anharmonicity of the lattice potential along the relevant phonon coordinate. This dual characterization reveals a dichotomy: Peierls-like distorted metavalent solids (Sb, GeTe, Te) exhibit both large deformation potentials and double-well lattice potentials, enabling the pronounced coherent phonon responses observed experimentally. By contrast, conventional covalent (GeSe, InSb) and semimetallic (AlPt) systems show small deformation potentials and nearly harmonic potentials, yielding rigid phonon dynamics. While fluence dependent phonon amplitude and softening have previously been reported for individual materials, our results show that these response functions are clearly correlated across bonding classes. This establishes a quantitative framework for classifying electron-lattice coupling and identifying materials suitable for ultrafast lattice control.

**Discussion**

This study establishes two complementary response functions, i.e., the increase of the coherent phonon amplitude and the pump-fluence-induced phonon softening rate, as quantitative metrics for comparing ultrafast lattice dynamics across a diverse set of solids. To ensure a meaningful comparison, both metrics are referenced to the absorbed pump fluence, calculated from measured optical constants, film thickness, and reflectance using the Beer-Lambert law. This normalization relates the observed lattice dynamics to the actual excitation density in each material rather than to the incident fluence.

Analysis of these two experimental response functions (Figure 3) reveals that materials in the conductivity regime between electron delocalization (metallic bonding) and electron localization (covalent bonding), such as Sb, GeTe, and $Bi_2Te_3$, exhibit both exceptionally large coherent A-mode amplitude-increase $S_A$ and pronounced fluence-induced frequency softening $S_\omega$. These materials have recently been classified as incipient metals that utilize an unconventional bonding mechanism coined metavalent bonding. [31, 37, 38] In contrast, covalent semiconductors, chiral topological semimetals, and layered transition-metal dichalcogenides display much weaker or negligible softening and substantially smaller amplitude-increase.

Part of the scatter in Fig. 3 may arise from variations in sample geometry, such as differences between bulk crystals and thin-film samples, which can influence optical detection efficiency. However, as demonstrated by the bulk to film comparisons in the Supplementary Information, these effects lead primarily to moderate variations in the absolute response coefficients and do not alter the qualitative trends across bonding classes.

The two experimental signatures probe complementary aspects of the coupled electron-lattice system. The amplitude increase reflects how strongly lattice motion along the phonon coordinate modulates the optical reflectivity $\Delta R/R$, which depends on both the efficiency with which the coherent phonon is generated and the sensitivity of the optical response to the lattice coordinate. The softening rate, in contrast, quantifies the renormalization of the phonon frequency under increasing excitation density and therefore reflects changes in the effective lattice potential. The observation that both optical signatures are maximized in a narrow set of compounds suggests a common underlying mechanism governing their sensitivity to ultrafast excitation. A simple upper-bound estimate of the transient lattice heating shows that the temperature increase expected for the applied pump fluences would lead to phonon frequency shifts substantially smaller than the softening observed here (see Supplementary Information). This indicates that the dominant contribution to the observed fluence-dependent softening arises from electronic excitation of the bonding network rather than from equilibrium lattice heating.

Quantum materials derive many of their unusual properties from competing electronic orders, i.e., correlation, topology, and lattice instabilities. While much attention focuses on correlation-driven phases or topologically protected states, metavalent bonding represents a lattice-dominated subgroup where Peierls-like localization competes directly with metallic delocalization without strong electronic correlations. In materials such as Sb, GeTe, $Bi_2Te_3$ and elemental Te, directional p-dominated bond networks realize a bonding state that is intermediate between covalent and metallic, with over-coordination, large bond polarizability, and an incipient Peierls-like distortion of a higher-symmetry parent structure. [31, 37, 39] The concept of metavalent bonding is closely related to the earlier notion of resonant bonding in crystalline phase-change materials, which emphasized the unusually high optical polarizability of these compounds. However, the term resonant bonding has historically been utilized to explain the structure and properties of aromatic molecules like benzene or solids like graphite. These materials neither feature a Peierls distortion, nor a pronounced anharmonicity. Hence, there is no reason to call phase change materials resonantly bonded, at least not if one strives for bonding concepts that help to understand and design material properties. The metavalent framework, on the contrary, provides a more quantitative classification. As such, it provides a scheme that distinguishes these incipient metals from both conventional covalent materials and good metals while retaining this historical connection. [40] The key advantage of the new bonding scheme is its predictive power.

Our frozen-phonon calculations indicate that this bonding environment produces two additional key microscopic features: large band-edge deformation potentials for fully symmetric A modes and strongly anharmonic, often double-well lattice potentials along the same coordinates. Within ground-state density functional theory, these calculations identify phonon coordinates that are particularly sensitive to electronic occupation near the Fermi level. Such modes are therefore especially susceptible to photoinduced force-constant renormalization under optical excitation. In this picture, photoexcited carriers can transiently shift the minimum of the lattice potential along the A-mode coordinate, producing large coherent displacements, while the reduced curvature of the potential yields strong fluence-dependent phonon softening. In contrast, more conventional covalent semiconductors (GeSe, InSb) and topological or transition-metal compounds with more metallic bonding (AlPt, $PtTe_2$) retain nearly harmonic A-mode potentials and smaller deformation potentials, consistent with their weak coherent amplitudes and negligible softening.

Additional support for this interpretation comes from thickness-dependent studies of metavalent materials. In previous work on Bi and GeTe thin films, we found that reducing the film thickness drives the system further into the distorted Peierls phase, leading to a systematic hardening of the corresponding A-mode phonon frequencies. [41-43] In other words, confinement moves the system away from the critical point, increasing the curvature of the lattice potential along the Peierls coordinate. Similar phonon hardening under reduced dimensionality has been reported for Sb and Te in the literature. [44-46] Together with the present fluence-induced softening results, this demonstrates that the same Peierls-like distortion coordinate in metavalent materials can be tuned in both directions, toward stronger distortion by structural confinement or toward reduced distortion by optical excitation.

This mechanism identifies metavalent materials as a distinct subgroup of quantum solids in which localization and delocalization are controlled primarily by the lattice through 3D Peierls-like distortions. This lattice-dominated electronic structure provides a microscopic explanation for why only these compounds combine unusually large coherent phonon amplitudes with strong pump-fluence-induced softening. In this sense, fully symmetric soft modes act as particularly effective handles for manipulating lattice structure and electronic properties on ultrafast timescales.

Within this framework, the two response functions introduced here provide practical experimental metrics for comparing electron-lattice coupling across material classes. While fluence-dependent phonon amplitude and softening have previously been reported for individual materials, our results demonstrate systematic trends across bonding regimes. The response functions therefore offer an operational way to identify materials with strongly enhanced lattice responses and to partition solids into groups with distinct ultrafast lattice dynamics.

Our comparative analysis is normalized to the optically determined absorbed fluence, which provides a consistent measure of excitation density across all samples. Although the response coefficients are extracted from slopes with respect to absorbed fluence, their absolute magnitude can still include the material-dependent optical detection factor $\partial R(\omega_{\mathrm{probe}})/\ \partial Q$. Because all samples are probed at the same wavelength, however, the response coefficients provide an experimentally consistent basis for comparing coherent phonon responses across materials. On the theoretical side, our frozen-phonon calculations employ ground-state density functional theory and harmonic phonons at zero temperature. Consequently, they neglect explicit anharmonicity, finite-temperature effects, and non-adiabatic electronic dynamics under strong excitation. Standard functionals also underestimate band gaps. These approximations limit quantitative accuracy under extreme nonequilibrium conditions but do not affect the central conclusion that metavalent systems possess significantly larger deformation potentials and more anharmonic lattice potentials than conventional covalent or metallic materials.

Our analysis focuses on Raman-active, fully symmetric phonons at the Brillouin-zone center, which dominate the coherent response in our optical pump-probe geometry. Materials lacking such Γ-point optical modes, such as *fcc* and *bcc* metals, or those where other symmetries or wave vectors are more relevant, will require adapted metrics based on different phonon branches or complementary probes. The general strategy of defining mode-specific response functions, however, is readily transferrable to those systems.

The present framework suggests several concrete directions for future work. Systematic studies with tunable pump and probe wavelengths could map how resonance conditions and selective excitation pathways influence coherent A-mode generation and softening across different material classes, while extending the same response functions to ultrafast electron and x-ray diffraction or time-resolved

ARPES would track the coupled evolution of lattice and electronic structure under non-equilibrium conditions. On the theoretical side, incorporating anharmonicity, finite-temperature effects, and explicit real-time electronic response through ab initio molecular dynamics or time-dependent electronic-structure methods will sharpen the connection to experiment and may uncover regimes of electron-lattice coupling beyond the harmonic, metavalent systems considered here.

Notably, our findings identify metavalent materials as particularly promising candidates in the search for hidden phases and ultrafast functional control. Their pronounced sensitivity to optical, thermal, electrical, and mechanical stimuli makes them especially amenable to phase manipulation and to the discovery of states that may be inaccessible in more rigidly bonded solids. Leveraging coherent phonon properties as a classification and screening tool may thus enable the targeted design of materials for ultrafast switching, phase manipulation, and other applications in emerging quantum technologies.

**Acknowledgments**

FH and MW acknowledge financial support from NeuroSys as part of the initiative “Clusters4Future”, which is funded by the Bundesministerium für Forschung, Technologie und Raumfahrt (BMFTR) (03ZU2106BA). MW acknowledges support from the Deutsche Forschungsgemeinschaft (DFG) for the transfer project within SFB 917. DMK acknowledges funding from the Deutsche Forschungsgemeinschaft (DFG, German Research Foundation) - 531215165 (Research Unit “OPTIMAL”). The authors gratefully acknowledge the computational the computing time provided to them at the NHR Center NHR4CES at RWTH Aachen University (project number p0022819). We thank N. Penner for valuable discussions on frozen-phonon calculations.

## Methods

*Femtosecond Pump-Probe Reflectivity*:

Ultrafast optical measurements were performed using a reflection-type, two-color pump-probe setup. The pump pulses were derived from a Ti:sapphire regenerative amplifier (central wavelength 800 nm, pulse duration ~60 fs, repetition rate 3 kHz). The pump beam was intensity-modulated with an optical chopper at 1.5 kHz and directed through a motorized optical delay line, allowing control of the pump-probe delay with femtosecond precision. The beam was focused onto the sample at near-normal incidence with a spot diameter of approximately 200 µm.

Probe pulses were generated by frequency conversion in an optical parametric amplifier followed by sum-frequency generation, producing pulses at 516 nm. The probe beam was focused onto the sample with a spot diameter of ~30 µm, ensuring homogeneous excitation within the probed region. Pulse durations were verified using intensity autocorrelation and frequency-resolved optical gating (FROG) measurements. Pump and probe beam spot sizes and fluences were determined using the Liu method from laser damage threshold measurements. Pump fluences were typically around 0.5-10 mJ/cm$^2$, while the probe fluence was kept approximately an order of magnitude lower to ensure perturbative readout.

The reflected probe beam was analyzed using polarization-resolved detection. A polarizing beam splitter separated the reflected probe into s- and p-polarized components that were detected by two balanced Si photodiodes connected to variable-gain current amplifiers. A third photodiode monitored the incident probe intensity for shot-to-shot normalization. This configuration enabled both isotropic transient reflectivity and anisotropic (electro-optic, EO) signals to be measured simultaneously.

The isotropic transient reflectivity was obtained from the summed signal of both polarization channels,

$$\Delta R/R = \frac{\Delta\left(R_s + R_p\right)}{R},$$

while the EO signal was derived from their difference,

$$\Delta R_{EO}/R = \frac{\Delta\left(R_s - R_p\right)}{R}.$$

To minimize systematic errors and slow experimental drifts, the order of delay positions was randomized for each scan. Pump-induced changes were obtained by comparing consecutive probe shots with the pump blocked and unblocked by the chopper. Reversibility of the optical excitation was verified by continuously monitoring the static reflectance of the sample during the measurements. All measurements were performed under ambient laboratory conditions.

*Density Functional Theory (DFT) Studies:*

To support the experimental findings and analyze the microscopic origin of the coherent phonon response, first principles calculations were performed within the framework of density functional theory (DFT). Structural relaxations, phonon calculations, and ground state energy calculations were carried out using the GGA PBE (Generalized Gradient Approximation by Perdew, Burke, and Ernzerhof) functional together with projector augmented wave (PAW) pseudopotentials as implemented in Quantum ESPRESSO 6.8 (QE) and VASP 5.4.4.

The primitive unit cells of the investigated bulk systems were first relaxed using "vc relax" calculations in QE, resulting in the following relaxed lattice constants: GeTe: a=b=c= 4.372 Å ,Sb: a=b=c= 4.583 Å, Te: a=b=4.506, c=5.055, GeSe: a=11.183 Å b=3.887 Å c=4.518 Å GaAs: a=b=c 4.065 Å ,AlPt: a=b=c = 4.917 Å. Subsequent frozen phonon calculations were performed by displacing atoms along the eigenvectors of the experimentally observed zone center phonon modes. The displacement amplitudes were chosen on the scale of the quantum zero-point motion of the respective modes to probe physically meaningful lattice distortions. For each displacement amplitude, total energies and electronic band structures were computed to determine the lattice potential energy surface and to extract deformation potentials of electronic states near the Fermi level.

The SCF calculations for the potential energy surface were conducted with a convergence threshold of 10^-10 Ry, no smearing was used for insulating systems. For metallic systems, a Gaussian smearing of 0.002 Ry was applied. k point grids were adapted to the crystal structure of each compound (AlPt: 20×20×20; GeSe: 8×23×20; GeTe: 20×20×20; InAs: 15×15×15; Sb: 15×15×15; Te: 15×15×15). An energy cutoff of 100 Ry was used.

To connect the lattice response with the experimentally measured optical properties, optical fingerprints, including the dielectric constant $\varepsilon_\infty$ and Born effective charges $Z^*$, were calculated as a function of frozen-phonon amplitude for GeSe and GeTe. These calculations were performed using the ph.x module within QE, which implements density functional perturbation theory (DFPT). The calculations were considered to be converged when the convergence threshold of 10-14 was achieved. For both systems no smearing was used. Additionally, VASP was used to compute the imaginary part of the dielectric function $\varepsilon_2(\omega)$, employing a 24×24×24 k-point grid, an energy cutoff of 550 eV, and a smearing of 0.1 eV. All calculations were conducted without spin-orbit coupling (SOC) or van-der-Waals (vdW) correction.

*Sample Preparation and Selection:*

Whenever available, bulk single crystals or polycrystalline samples were used for the optical pump–probe measurements. For materials where bulk crystals were not available, thin films were employed. In these cases, the thickest available films were selected in order to ensure sufficient optical absorption of the pump pulse.

Most thin-film samples were grown by molecular beam epitaxy (MBE) and protected by a thin $Al_2O_3$ capping layer to prevent surface oxidation. Film thicknesses were chosen such that the optical penetration depth of the pump pulse was well contained within the film. The film thickness was also explicitly considered when calculating the absorbed pump fluence using the Beer–Lambert law.

For a small number of compounds where MBE-grown samples were not available, polycrystalline thin films prepared by magnetron sputtering were used. All samples were characterized structurally prior to the optical measurements to verify phase purity and crystallinity.

Although absolute coherent phonon amplitudes and softening rates can be influenced by sample geometry, crystallographic defects, and optical boundary conditions in thin-film structures, the normalization to absorbed pump fluence and the comparison across multiple bonding classes reveal robust trends that are unlikely to arise solely from sample-specific effects. A comparison of bulk and thin-film samples for representative materials is provided in the Supplementary Information.

**Supplementary Information – Divergent Coherent Phonon Responses Across the Metal-Insulator Crossover**

*Felix Hoff, Timo Veslin, Tim Bartsch, Carl-Friedrich Schön, Dante M. Kennes, Matthias Wuttig*

F. Hoff, T. Veslin, T. Bartsch, C-F. Schön, M. Wuttig

Institute of Physics IA, RWTH Aachen University, 52074 Aachen, Germany

E-Mail: wuttig@physik.rwth-aachen.de

D. M. Kennes

Institute for Theory of Statistical Physics, RWTH Aachen University, and JARA Fundamentals of Future Information Technology, 52074 Aachen, Germany

D. M. Kennes

Max Planck Institute for the Structure and Dynamics of Matter, 22761 Hamburg, Germany

M. Wuttig

Peter-Grünberg-Institute – JARA-Institute Energy Efficient Information Technology (PGI-10), 52428 Jülich, Germany

The Supplementary Information provides additional methodological details and supporting analyses that complement the results presented in the main text. Specifically, we document the procedures used to determine the two experimental response functions, the coherent phonon amplitude-increase $\Delta A$ and the phonon softening rate $\Delta\omega/\omega$, from the pump-probe data. We also describe the normalization of the excitation density using the absorbed pump fluence derived from the Beer-Lambert law and the optical constants of the investigated materials. Furthermore, we provide additional information on the frozen-phonon calculations, including the extraction of band-edge deformation potentials. Finally, we demonstrate the robustness of the amplitude analysis by comparing several alternative figures of merit for oscillation strength and showing their strong mutual correlation. These supplementary analyses confirm that the trends reported in the main text are independent of the specific metric used and reflect intrinsic material properties.

*Extraction of the coherent phonon response functions*

Figure S1 illustrates the procedure used to extract the two response functions discussed in the main text: the amplitude-increase and the phonon softening rate. The example shown for $Bi_2Te_3$ demonstrates the analysis workflow from raw transient reflectivity traces to the linear fluence dependences used to determine both quantities.

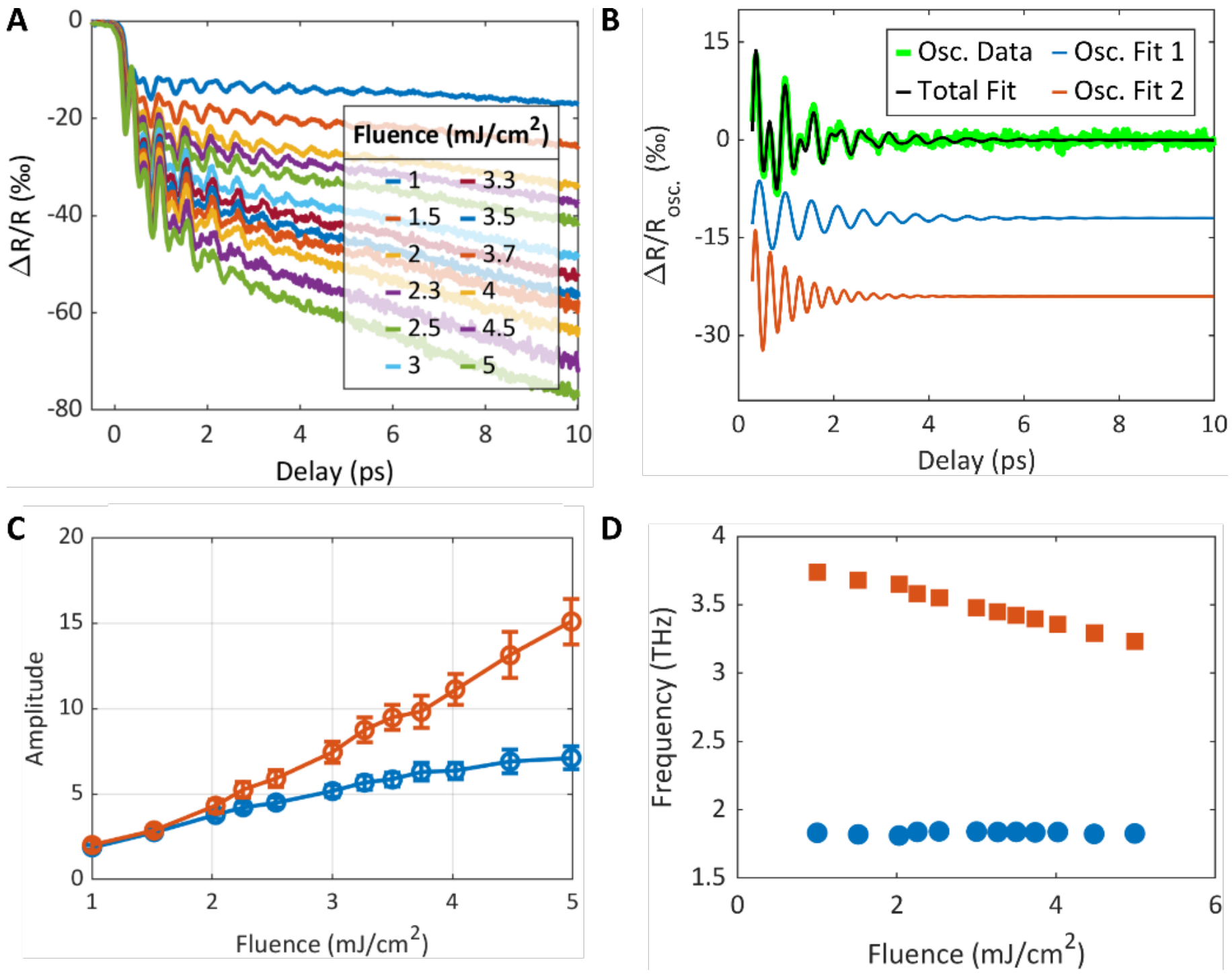

**Figure S1**: Determination of the two experimental response functions for $Bi_2Te_3$. (**A**) Raw transient reflectivity traces $\Delta R/R$ as a function of pump-probe delay for a series of incident pump fluences. (**B**) Example decomposition of the oscillatory component obtained after subtraction of the non-oscillatory background. The residual oscillations are fitted with a sum of damped harmonic oscillator (DHO) functions (individual mode fits and total fit shown), from which the phonon amplitudes and frequencies are extracted. (**C**) Extracted coherent phonon amplitudes (sum of fitted oscillator amplitudes) as a function of incident pump fluence. The slope of a linear fit defines the amplitude-fluence response coefficient used in the main text. (**D**) Corresponding phonon frequencies as a function of incident pump fluence. A linear fit yields the fluence-induced phonon softening rate. All panels are shown as a function of incident pump fluence; the conversion to absorbed fluence used in the main analysis is described in Fig. S2.

*Determination of Absorbed Pump Fluence*

To enable a consistent comparison of excitation density across materials, the pump fluence used in the analysis is expressed as the absorbed fluence rather than the incident fluence. The absorbed fraction of the incident light is calculated using the Beer–Lambert law together with the optical constants of the material.

The intensity inside the film decays according to

$$I(z) = I_0 e^{-\alpha z},$$

where $z$ is the depth and $\alpha$ is the absorption coefficient. The latter is related to the extinction coefficient $k$ through

$$\alpha = \frac{4\pi k}{\lambda}.$$

For a film of thickness $d$, the absorbed fraction of the incident fluence is

$$A = (1 - R)\left(1 - e^{-\alpha d}\right),$$

where $R$ is the reflectance at the surface. This expression corresponds to the Beer–Lambert approximation for optical absorption. For thin films with capping layers or substrates, interference and field-distribution effects can in principle modify the exact absorbed fraction. However, the films investigated here were chosen such that the optical penetration depth of the pump pulse is largely contained within the film thickness, and the Beer–Lambert estimate therefore provides a reliable first-order approximation of the absorbed excitation density used for the comparative analysis.

The extinction coefficient $k$ is obtained from the dielectric function via

$$|\varepsilon| = \sqrt{\varepsilon_1^2 + \varepsilon_2^2}, \qquad k = \sqrt{\frac{|\varepsilon| - \varepsilon_1}{2}}.$$

The absorbed pump fluence used in the main analysis is then

$$F_{\mathrm{abs}} = A \cdot F_{\mathrm{inc}}.$$

Figure S2 illustrates this procedure for $Bi_2Te_3$.

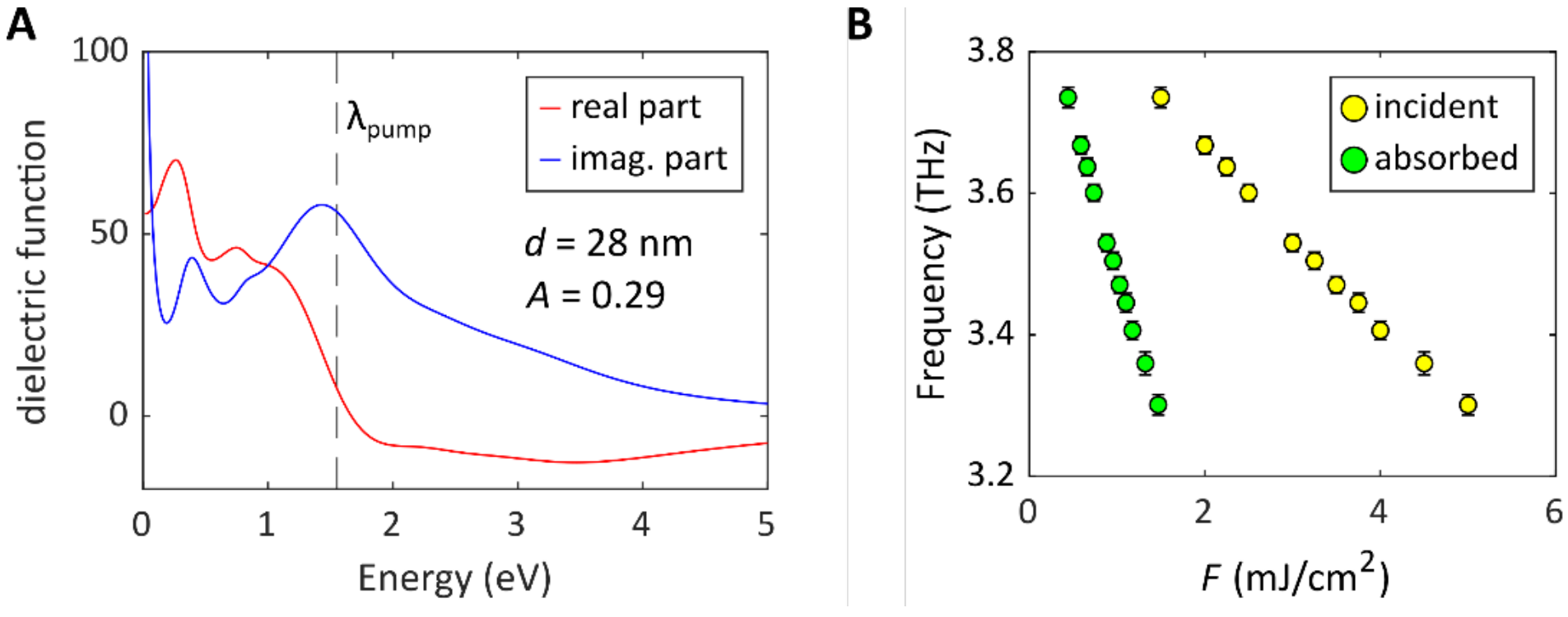

**Figure S2**: Determination of absorbed pump fluence for Bi2Te3. (**A**) Real and imaginary parts of the dielectric function ($\varepsilon_1$, $\varepsilon_2$). The vertical line indicates the pump wavelength (800 nm) used in the experiments. (**B**) Example

of the phonon frequency plotted as a function of absorbed pump fluence after applying the normalization described in the text.

*Statistical Uncertainty of the Extracted Response Functions*

To estimate the statistical uncertainty of the extracted response functions, the analysis was performed on the level of individual pump-probe traces. For each transient reflectivity trace, the non-oscillatory background was fitted and subtracted, and the remaining oscillatory component was fitted with a damped harmonic oscillator function, yielding independent estimates of the coherent phonon amplitude and frequency for each repetition at a given pump fluence.

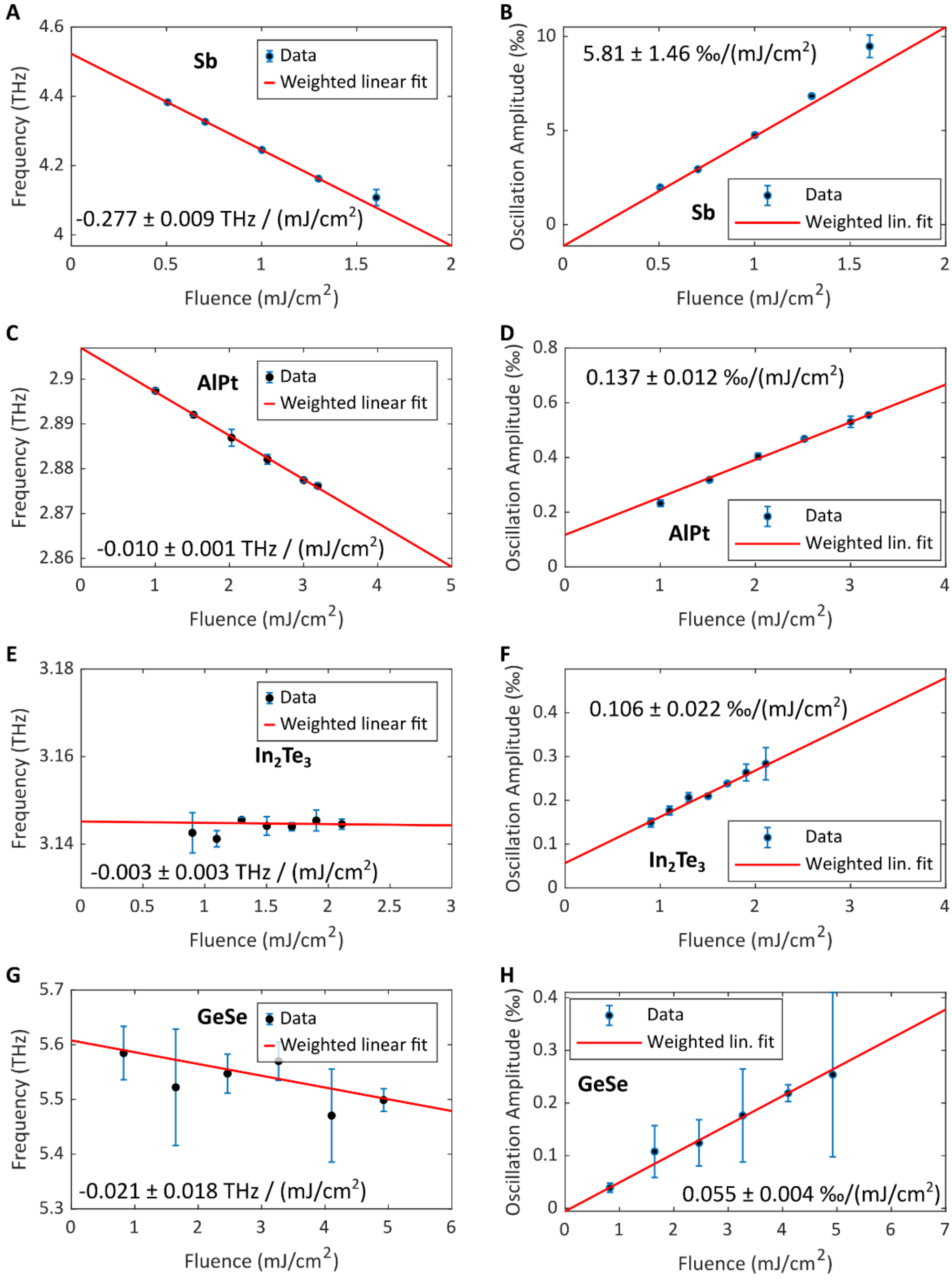

Figure S3: Incident fluence dependence of coherent phonon amplitude and frequency for representative materials from different bonding classes. Symbols show the mean values obtained from repeated measurements

at each fluence, with error bars representing the standard deviation. Solid lines indicate weighted linear fits used to extract the response coefficients reported in the main text. Slopes and intercepts with their uncertainties (95% confidence interval) are indicated in each panel.

For every fluence value, the reported parameter corresponds to the arithmetic mean of the values obtained from all repetitions, while the uncertainty is given by the standard deviation, reflecting the shot-to-shot variability of the experiment and the sensitivity of the fitting procedure to measurement noise.

The fluence dependence was determined by fitting the averaged quantities with a weighted linear regression. The weights were chosen as the inverse variance of the measured values. The uncertainties of the slope and intercept were obtained from the 95% confidence interval of the weighted regression. Because the weighting is based on the statistical spread of independently fitted repetitions, the resulting confidence intervals provide a realistic estimate of the uncertainty of the extracted fluence dependence.

For each material, the linear regression was restricted to the reversible low-to-intermediate fluence regime where the coherent phonon amplitude and frequency exhibit approximately linear behavior with excitation density. Data points at higher fluence showing clear deviations from linearity, such as amplitude saturation or strong frequency chirping, were excluded from the fit.

*Thermal Contributions to Phonon Softening*

To estimate the possible influence of transient lattice heating, we performed an upper-bound estimate of the temperature rise using

$$\Delta T = \frac{F_{abs}}{d_{eff} \rho C_p}.$$

For GeTe, using $F_{abs} = 1\,\mathrm{mJ/cm^2}$, an effective absorption depth $d_{eff} = 17.2\,\mathrm{nm}$, density $\rho = 6.14\,\mathrm{g/cm^3}$, and heat capacity $C_p = 0.25\,\mathrm{J/gK}$, this estimate yields $\Delta T \approx 380\,\mathrm{K}$ per $1\,\mathrm{mJ/cm^2}$.

The equilibrium Raman temperature coefficient of the GeTe $A_1$ phonon is approximately $d\nu/dT \approx -0.0186\,\mathrm{cm^{-1}/K}$, corresponding to an estimated thermal phonon shift of about $\Delta\nu \approx -7\,\mathrm{cm^{-1}}$, or roughly -5% of the phonon frequency per $1\,\mathrm{mJ/cm^2}$. This value represents a conservative upper bound because it assumes instantaneous conversion of the absorbed optical energy into lattice heat and neglects the finite electron-phonon equilibration time.

Experimentally, however, the phonon softening observed for GeTe reaches values of $\Delta\omega/\omega \approx -30\%$ per $1\,\mathrm{mJ/cm^2}$, far exceeding the shift expected from equilibrium heating alone. In addition, the large fluence-dependent softening occurs predominantly in metavalent materials, while covalent semiconductors and metallic systems show much weaker effects under comparable excitation conditions. This strong material selectivity further indicates that the dominant mechanism is electronic in origin, consistent with photoexcitation-induced weakening of the Peierls-like bonding network rather than simple lattice heating.

*Deformation Potentials of Electronic States Near the Fermi level*

Deformation potentials were extracted from the frozen-phonon calculations by tracking the energy shifts of electronic states along the relevant high-symmetry k-paths. For semiconducting systems such as GeTe and GeSe, this analysis focuses on the valence- and conduction-band extrema. For semimetallic or metallic systems, the analysis instead considers electronic states in the vicinity of the Fermi level that contribute most strongly to the optical and transport properties.

The energies of the valence-band maximum (VBM) and conduction-band minimum (CBM) were tracked as a function of the phonon amplitude $Q$. The deformation potential $D$ of a given band is defined as the slope

$$D = \frac{\partial E}{\partial Q},$$

obtained from a linear fit of the band energy $E(Q)$ with respect to the displacement amplitude $Q$. The deformation potentials were evaluated at the relevant band extrema of each material (e.g., at the $Z$ point for GeTe).

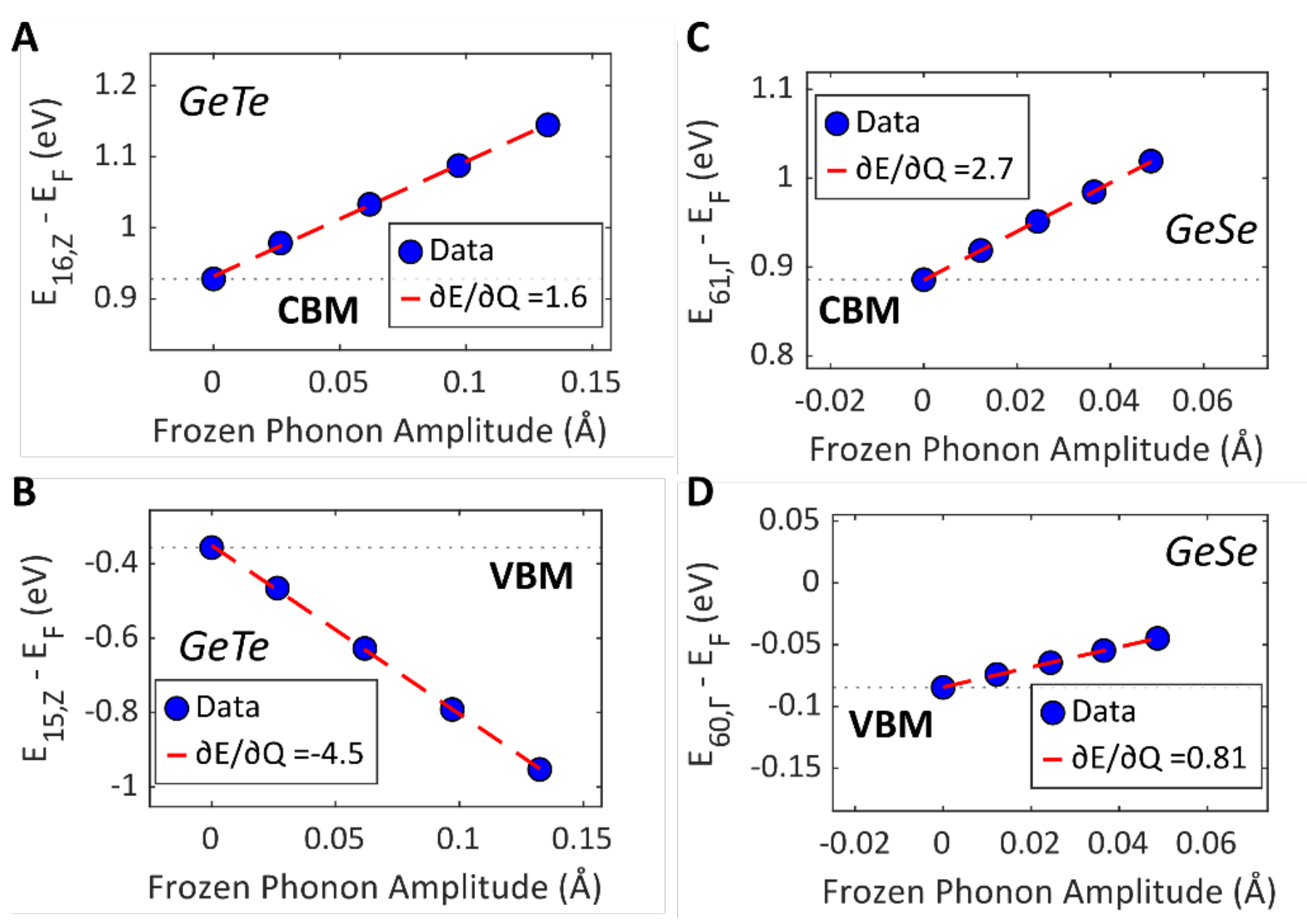

**Figure S4**: Extraction of band-edge deformation potentials from frozen-phonon calculations. (**A,B**) Energies of the valence-band maximum (VBM) and conduction-band minimum (CBM) of GeTe at the $Z$ point as a function of the frozen-phonon displacement amplitude $Q$. Linear fits yield deformation potentials of $-4.5\,\mathrm{eV/Å}$ (VBM) and $+1.6\,\mathrm{eV/Å}$ (CBM). (**C,D**) Corresponding analysis for GeSe at the band extrema. The extracted deformation potentials are $+0.81\,\mathrm{eV/Å}$ (VBM) and $+2.7\ \mathrm{eV/Å}$ (CBM). The opposite signs of the slopes in GeTe reflect the modulation of bonding-antibonding splitting along the Peierls distortion coordinate, whereas the same-sign slopes in GeSe indicate a much weaker coupling of the phonon mode to the band-edge electronic states.

Figure S3 illustrates this procedure for GeTe and GeSe. In GeTe, the CBM and VBM exhibit deformation potentials of $+1.6\ \mathrm{eV/Å}$ and $-4.5\ \mathrm{eV/Å}$, respectively, indicating that the frozen-phonon displacement directly modulates the bonding-antibonding splitting of the $p$-derived electronic states along the Peierls axis. In contrast, GeSe shows deformation potentials of $+2.7\ \mathrm{eV/Å}$ (CBM) and $+0.81\ \mathrm{eV/Å}$ (VBM), with both bands shifting in the same direction. This behavior reflects the fact that the excited phonon mode in GeSe does not strongly tune a Peierls-type bonding coordinate, consistent with the weaker coherent phonon response observed experimentally.

Table S1 summarizes the materials investigated in this study together with key information relevant for the comparative analysis presented in the main text. For each material we list the crystal structure (space group), the dominant coherent phonon mode observed in the pump-probe measurements, the sample type (thin film or bulk (single) crystal), the approximate room-temperature electrical conductivity, and the bonding classification used for the color coding in Fig. 3. The bonding assignment follows the multi-property classification framework discussed in the main text, which distinguishes covalent solids, metavalent materials (incipient metals), and topological chiral semimetals based on their bonding characteristics and electronic properties. [21,27,28] The TCSM data originate from measurements reported elsewhere (manuscript under review). The listed conductivities provide an approximate indicator of the electronic transport regime and are included to illustrate the broad range of conductivities covered in this work.

Table 1: Overview of the materials included in the comparative coherent-phonon study. Listed are the crystal structure, the dominant zone-center phonon mode observed in the pump-probe measurements, the sample type, approximate room-temperature electrical conductivity (in S/cm), and the bonding classification used in Fig. 3.

| Material | Structure | Dominant Phonon | Sample type | Conductivity (S/cm) | Bonding Class |
|---|---|---|---|---|---|
| Sb | R-3m | A1g | Film | 2.50E+04 | Metavalent |
| Bi | R-3m | A1g | Film | 7.70E+03 | Metavalent |
| BiTe | P-3m1 | A1g | Film | 3.67E+03 | Metavalent |
| Bi2Te | P-3m1 | A1g | Film | 1.00E+03 | Metavalent |
| Bi4Te3 | R-3m | A1g | Film | 8.00E+02 | Metavalent |
| Bi7Te3 | R-3m | A1g | Film | 1.00E+02 | Metavalent |
| Bi9Te10 | R-3m | A1g | Film | 3.70E+03 | Metavalent |
| Bi2Te3 | R-3m | A_1g^2 | Film | 6.60E+02 | Metavalent |
| Sb2Te3 | R-3m | A_1g^1 | Film | 2.30E+03 | Metavalent |
| Sb2Te | P-3m1 | A1g | Film | 1.50E+03 | Metavalent |
| Bi2Se3 | R-3m | A_1g^2 | Bulk Crystal | 1.00E+03 | Metavalent |
| Te | P3_121 | A1g | Bulk Crystal | 1.00E+02 | Metavalent |
| GST124 | R3m | A1 | Film | 3.50E+02 | Metavalent |
| GeTe | R3m | A1 | Film | 5.00E+03 | Metavalent |
| In2Te3 | F-43m | A1g | Film | 1.25E-01 | Covalent |
| InTe | I4/mcm | A1g | Film | 1.00E+01 | Covalent |
| Sb2Se3 | Pnma | Ag | Film | 4.00E-07 | Covalent |
| Sb2S3 | Pnma | Ag | Film | 1.00E-08 | Covalent |
| SnSe | Pnma | Ag | Bulk Crystal | 1.00E+01 | Covalent |
| GeSe | Pnma | Ag | Film | 1.30E-06 | Covalent |
| InSb | F-43m | LO (T_2) | Bulk Crystal | 2.20E+02 | Covalent |
| PdSe2 | Pbca | A1g | Bulk Crystal | n/a | Covalent |
| SnS | Pnma | Ag | Bulk Crystal | 1.30E-01 | Covalent |
| InAs | F-43m | TO (T_2) | Bulk Crystal | 2.50E+02 | Covalent |
| GaPt | P2_13 | A | Bulk Crystal | 4.54E+04 | TCSM |
| AlPt | P2_13 | A | Bulk Crystal | 1.96E+04 | TCSM |
| GaPd | P2_13 | A | Bulk Crystal | 4.54E+04 | TCSM |
| RhSi | P2_13 | A | Bulk Crystal | 5.70E+03 | TCSM |
| Zn [lit.] | P6_3/mmc | E_2g | Bulk Crystal | 1.70E+05 | Metal |

*Comparing Bulk and Film Samples*

To evaluate the influence of sample geometry on the extracted response functions, we compared coherent phonon measurements obtained from both bulk and thin-film samples for two representative materials: the metavalent compound GeTe and the covalent semiconductor SnSe. For GeTe, measurements were performed on a bulk polycrystalline sample and a sputtered thin film (300 nm on Si with ZnS-$SiO_2$ capping). For SnSe, we compared a bulk polycrystalline crystal with an epitaxial thin film grown by molecular beam epitaxy (75 nm on Si with $Al_2O_3$ capping).

All datasets were analyzed using the same procedure described above, extracting coherent phonon amplitudes and frequencies from individual pump-probe traces and determining the response coefficients from linear fits within the reversible excitation regime.

Figure S6 compares the fluence dependence of the coherent phonon amplitude and frequency for bulk and thin-film samples of both materials. While the absolute slopes differ between bulk and thin-film geometries, the qualitative behavior remains consistent. GeTe exhibits a pronounced increase of coherent phonon amplitude and strong fluence-dependent phonon softening for both sample forms, characteristic of metavalent materials. In contrast, SnSe shows only very weak amplitude growth and minimal phonon softening in both bulk and thin-film samples, consistent with the behavior expected for conventional covalent semiconductors.

The quantitative differences between bulk and thin-film slopes can arise from variations in strain, defect density, and optical boundary conditions in thin-film geometries. In particular, the detection efficiency of coherent phonons in optical reflectivity experiments depends on the derivative $\partial R/\partial Q$, which can be modified by the sample geometry. Nevertheless, the comparison demonstrates that the qualitative trends of the response functions are robust across different sample forms, supporting the conclusion that the large coherent phonon response observed in Fig. 3 is primarily governed by intrinsic bonding characteristics rather than by sample geometry.

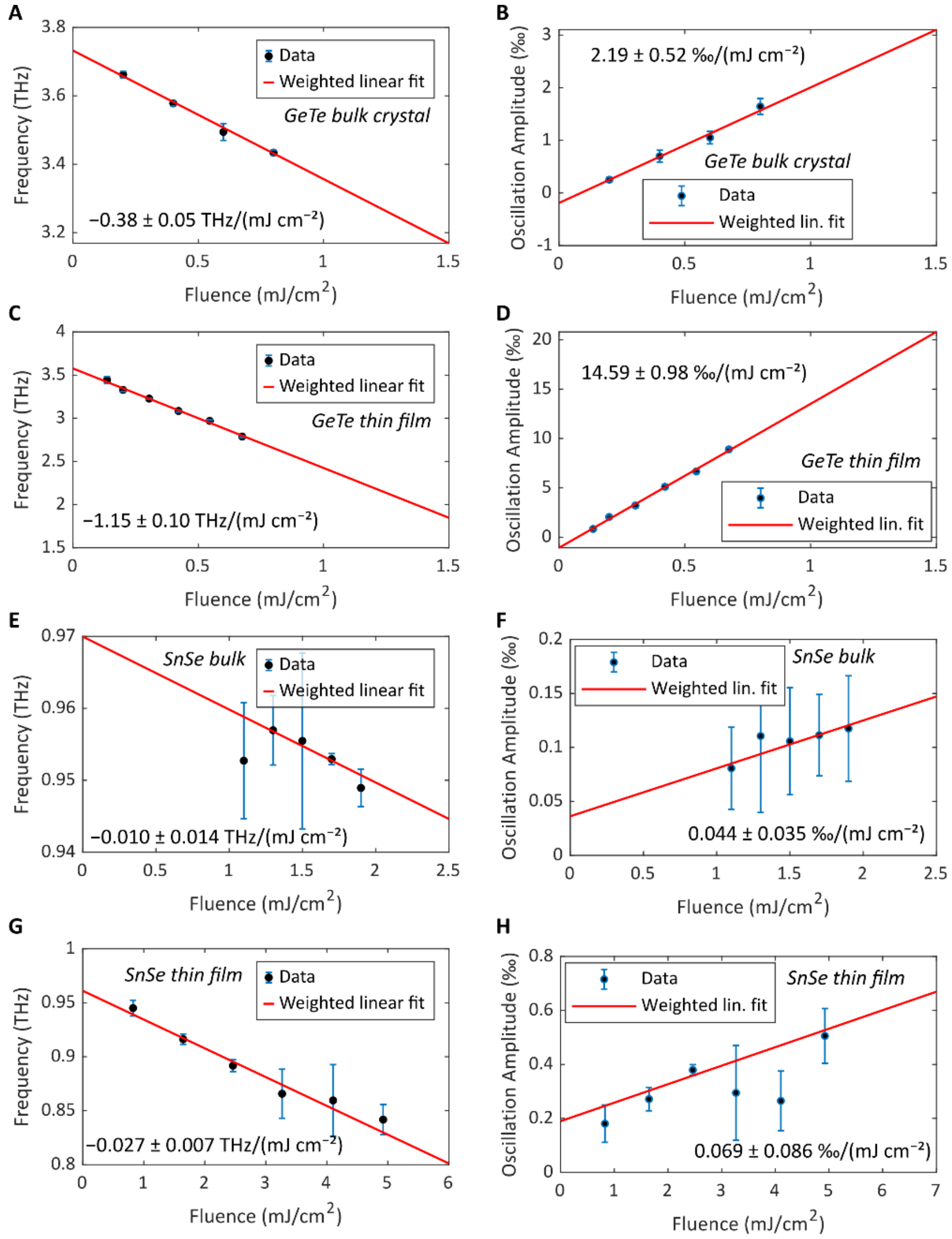

Figure S5: Comparison of coherent phonon response functions for bulk and thin-film samples. Panels (A,B) show the incident fluence dependence of the coherent phonon amplitude and frequency for a bulk polycrystalline GeTe sample, while (C,D) show the corresponding data for a sputtered GeTe thin film (300 nm on Si with ZnS-$SiO_2$ capping). Panels (E,F) display measurements for a bulk polycrystalline SnSe crystal, and (G,H) for an epitaxial SnSe thin film (75 nm grown by MBE). Symbols represent the extracted phonon amplitudes and frequencies as a function of incident pump fluence, and solid lines indicate linear fits used to determine the response coefficients. While the absolute slopes vary between bulk and thin-film samples due to differences in crystallographic defects, and optical boundary conditions, the qualitative behavior remains consistent: GeTe exhibits strong amplitude growth and pronounced phonon softening characteristic of metavalent materials, whereas SnSe shows only weak amplitude changes and minimal softening typical of covalent semiconductors.

*Robustness of amplitude metrics*

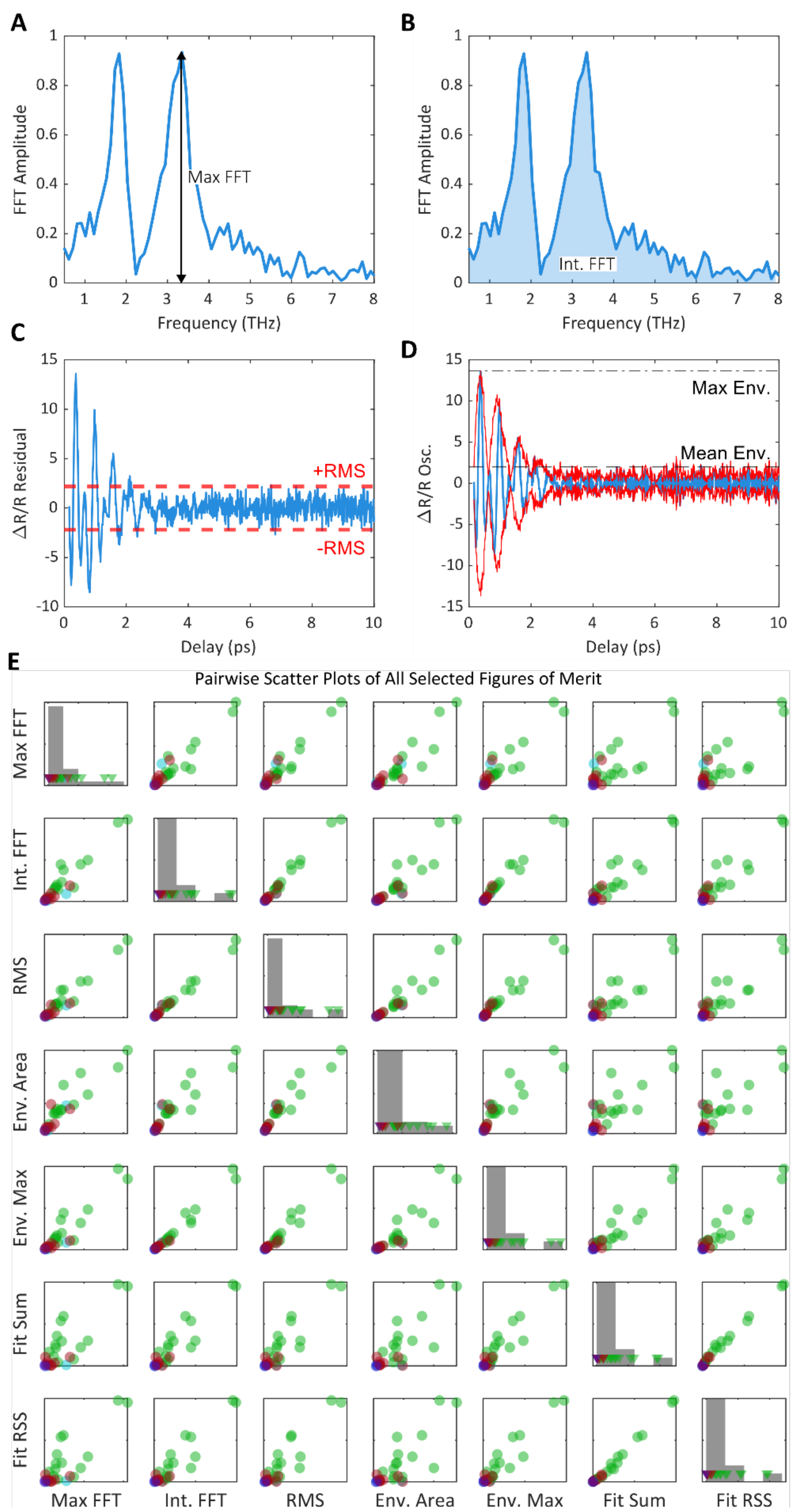

**Figure S6**: Robustness of coherent phonon amplitude metrics. (**A-D**) Comparison of several figures of merit used to quantify oscillation strength for $Bi_2Te_3$: maximum FFT amplitude, integrated FFT amplitude, RMS amplitude of the residual signal, and maximum and mean envelope values. (**E**) Pairwise scatter plots of the different amplitude metrics evaluated across the dataset, showing strong agreement between the methods.

To verify that the trends reported in the main text do not depend on the specific metric used to quantify oscillation strength, several alternative figures of merit were evaluated for the transient reflectivity data. These include frequency-domain measures such as the maximum FFT peak and integrated FFT amplitude, time-domain measures such as the root-mean-square (RMS) amplitude of the oscillatory residual, and amplitudes extracted from damped harmonic oscillator fits. Figure S4 illustrates the comparison for a representative dataset ($Bi_2Te_3$). All metrics are strongly correlated across the dataset. This confirms that the trends discussed in the main text are robust and independent of the specific definition of oscillation strength.